\documentclass[12pt]{iopart}
\pdfoutput=1

\expandafter\let\csname equation*\endcsname\relax
\expandafter\let\csname endequation*\endcsname\relax

\usepackage{amsmath,amsfonts,amssymb}
\usepackage[english]{babel} 
\usepackage[latin1]{inputenc} 
\usepackage[T1]{fontenc}
\usepackage{color}
\usepackage{float}
\usepackage{verbatim}
\usepackage{graphicx}
\usepackage{bm}
\usepackage{mathtools}
\usepackage{subfigure}
\usepackage{cite}

\newcommand{\partialderiv}[2]{\frac{\partial #1}{\partial #2}}
\newcommand{\diff}[1]{\mathrm{d}#1 \;}

\begin{document}

\title{Optimal strategy to capture a skittish lamb wandering near a
  precipice}

\author{M. Chupeau}
\address{Laboratoire de Physique Th\'eorique de la Mati\`ere Condens\'ee (UMR CNRS 7600), 
Universit\'e Pierre et Marie Curie, 4 Place Jussieu, 75255 Paris Cedex France}

\author{O. B\'enichou}
\address{Laboratoire de Physique Th\'eorique de la Mati\`ere 
Condens\'ee (UMR CNRS 7600), Universit\'e Pierre et Marie Curie, 4 
Place Jussieu, 75255 Paris Cedex France}

\author{S. Redner} 
\address{Santa Fe Institute, 1399 Hyde Park Road, Santa Fe, NM 87501, USA}
\address{Center for Polymer Studies and Department of Physics, Boston
  University, Boston, MA 02215, USA}

\begin{abstract}
  We study the splitting probabilities for a one-dimensional Brownian motion
  in a cage whose two boundaries move at constant speeds $c_1$ and $c_2$.
  This configuration corresponds to the capture of a diffusing, but skittish
  lamb, with an approaching shepherd on the left and a precipice on the
  right.  We derive compact expressions for these splitting probabilities
  when the cage is expanding.  We also obtain the time-dependent
  first-passage probability to the left boundary, as well as the splitting
  probability to this boundary, when the cage is either expanding or
  contracting.  The boundary motions have a non-trivial impact on the
  splitting probabilities, leading to multiple regimes of behavior that
  depend on the expansion or contraction speed of the cage.  In particular,
  the probability to capture the lamb is maximized when the shepherd moves at
  a non-zero optimal speed if the initial lamb position and the ratio between
  the two boundary speeds satisfy certain conditions.
\end{abstract}
\pacs{02.50.Ey, 05.40.Jc}
\maketitle

\section{Introduction}

A lamb escapes from a farm and has the bad idea to roam near a precipice.
The shepherd wonders what is the best strategy to catch her lamb alive.
Indeed, the lamb wanders randomly when the shepherd stays still.  However the
lamb is skittish and moves away from the shepherd and toward the precipice
whenever the shepherd approaches.  Should the shepherd stay still and hope
the lamb will come to her, or should she walk toward the lamb and hope that
she reaches the lamb before it goes over the precipice?

\begin{figure}[ht]
\centering
\includegraphics[width=300pt]{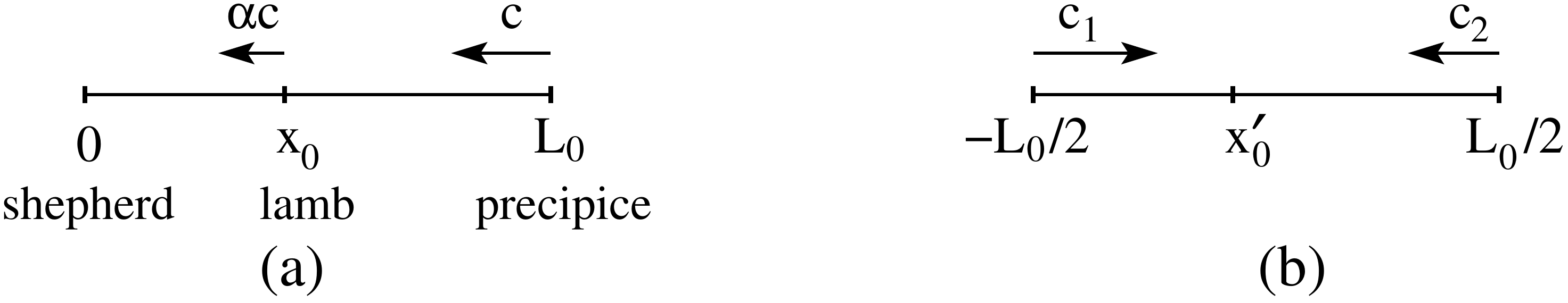}
\caption{Equivalent formulations of the lamb capture process.  (a) Shepherd's
  reference frame: the lamb and the precipice drift toward the shepherd at
  speeds $\alpha c$ and $c$.  (b) Reference frame where the lamb only diffuses: the shepherd and
  the precipice approach the lamb at speeds $c_1=\alpha c$ and
  $c_2=(1-\alpha) c$.}
\label{setup}
\end{figure}

If the shepherd walks toward the lamb with speed $c$, we assume that the lamb
moves away at a slower speed $(1-\alpha)c$, with
$0\leqslant \alpha \leqslant 1$, that is superimposed on its diffusive
motion.  In the reference frame of the shepherd (Fig.~\ref{setup}(a)), the
lamb diffuses and approaches the shepherd with speed $\alpha c$.  However,
the precipice also approaches the lamb with speed $(1-\alpha)c$.  Initially,
the lamb is at $x_0$, the precipice is at $L_0$, while the shepherd is fixed
at the origin.  The probability that the lamb and shepherd meet before the
lamb goes over the precipice coincides with the \emph{splitting probability}
for the lamb to eventually reach the origin.  In turn, this latter problem is
equivalent to the splitting probability for unbiased diffusion that starts at
$x_0'=x_0\!-\!L_0/2$ to reach the left edge of an asymmetrically contracting
cage, whose left and right edges are at $-L_0/2\!+\!c_1 t$ and
$L_0/2\!-\!c_2 t$, respectively, with $c_1=\alpha c$ and
$c_2=(1\!-\!\alpha) c$ (Fig.~\ref{setup}(b)).  While the contracting cage is
the relevant situation for the lamb capture problem, we also investigate the
case of the expanding cage.  Although the value of $\alpha$ does not need to
be restricted, we study the range $\alpha \in [0,1]$, where the splitting
probability has the richest behavior.

Conditional exit from a fixed interval is a classic problem of random-walk
theory \cite{Gardiner,VanKampen,Weiss,Redner}. The role of a moving boundary
has been considered recently, e.g., a diffusing or an oscillating trap at the
edge of the interval \cite{Tzou:2014,Holcman:2009,Tejedor:2011}.  For ballistically moving
boundaries, which is our focus, the infinite-time survival probability in an
asymmetrically expanding cage and the time-dependent survival probability in
a symmetrically expanding cage were investigated by Bray and
Smith~\cite{Bray:2007,Bray:2007b}. Here we derive exact expressions for the
first-passage probability at any time and the splitting probabilities to one
of the edges of the cage when each wall moves ballistically at an arbitrary
speed. Contracting and expanding cages are considered. The surprising consequence of our results is that the splitting
probabilities depend \emph{non-monotonically} on the speed $c$ for a range of
initial conditions and $\alpha$ values.

In Sec.~\ref{sec:exp}, we derive the splitting probability in an expanding
cage.  In Sec.~\ref{sec:gen}, we determine the first-passage probability at
any time and the splitting probability to one of the boundaries for both an
expanding and contracting cage.  Finally, in Sec.~\ref{sec:disc}, we
demonstrate that the splitting probability can have a non-monotonic
dependence on the speed $c$.  Using these results, we answer the shepherd's
question of what is the best strategy to capture the lamb.

\section{ Expanding Cage}
\label{sec:exp}

It is convenient to treat the problem in the reference frame of the left
boundary (Fig.~\ref{setup}(a)), where the Brownian particle drifts to the
right with speed $\alpha c$ and the right boundary also drifts to the right
with speed $c$.  Thus the right boundary is located at $L(t)=L_0\!+\!ct>L_0$.
We focus on the splitting probability to reach the left edge of an expanding cage
$\mathcal{L}^{\mathrm{e}}(x_0,L_0)$ as a function of the initial particle
position $x_0$ and the initial interval length $L_0$ (with
$\mathcal{R}^{\mathrm{e}}$ the splitting probability to the right
edge).  Following the approach of Ref.~\cite{Bray:2007}, the backward
Fokker-Planck equation for the splitting probability is
\begin{subequations}
\begin{equation}
D \frac{\partial^2\mathcal{L}^{\mathrm{e}}}{\partial x_0^2} +\alpha c \frac{\partial
 \mathcal{L}^{\mathrm{e}}}{\partial x_0}+c \frac{\partial\mathcal{L}^{\mathrm{e}}}{\partial L_0}=0\,, 
\end{equation}
or, by introducing the rescaled variables $y=c x_0/D$ and $\lambda=cL_0/D$, 
\begin{equation}\label{FPasym}
\frac{\partial^2 \mathcal{L}^{\mathrm{e}}}{\partial y^2} +\alpha \partialderiv{\mathcal{L}^{\mathrm{e}}}{y} +\partialderiv{\mathcal{L}^{\mathrm{e}}}{\lambda}=0\,, 
\end{equation}
\end{subequations}
with $0 \leqslant y \leqslant \lambda$ and the boundary conditions
$\mathcal{L}^{\mathrm{e}}(0,\lambda)=1$ and $\mathcal{L}^{\mathrm{e}}(\lambda,\lambda)=0$. 

In the spirit of \cite{Bray:2007}, we seek a solution of the form
\begin{equation}
\label{gen}
\mathcal{L}^{\mathrm{e}}(y,\lambda)=\sum\limits_{n\in \mathbb{Z}} \big[a_n e^{ny} \!+\! b_n e^{-(n+\alpha)y} \big] e^{-(n+\alpha)n\lambda}.
\end{equation}
The left boundary condition $\mathcal{L}^{\mathrm{e}}(0,\lambda)=1$ leads to
\begin{subequations}
\label{CL}
\begin{equation}\label{CLleft}
\mathcal{L}^{\mathrm{e}}(0,\lambda)=\sum\limits_{n\in\mathbb{Z}} (a_n+b_n) \mathrm{e}^{-n(n+\alpha)\lambda}=1\,,
\end{equation}
while the right boundary condition $\mathcal{L}^{\mathrm{e}}(\lambda,\lambda)=0$ gives
\begin{equation*}
\label{bc-right}
\mathcal{L}^{\mathrm{e}}(\lambda,\lambda)=\sum\limits_{n\in\mathbb{Z}} 
\big[a_n e^{n\lambda}+b_ne^{-(n+\alpha)\lambda}\big] e^{-(n+\alpha)n\lambda}=0\,.
\end{equation*}
By shifting the index of the second sum, $n\to n-1$, and after some simple
algebra, the condition above can be written as
\begin{equation}\label{CLright}
\mathcal{L}^{\mathrm{e}}(\lambda,\lambda)=\sum\limits_{n\in\mathbb{Z}} (a_n+b_{n-1}) \mathrm{e}^{n(1-n-\alpha)\lambda}=0\,.
\end{equation}
\end{subequations}
Since Eqs.~\eqref{CL} hold for all $\lambda$, we obtain the following
relations for the coefficients in \eqref{gen}:
\begin{eqnarray}\label{relations}
\begin{cases}
a_0+b_0=1\,, \\
a_n+b_n=0 &\qquad \forall n \neq 0\,, \\
a_n+b_{n-1}=0 &\qquad \forall n\,. \\
\end{cases}
\end{eqnarray}

If the initial cage length $L_0\to\infty$ and the initial position of the
particle is far from either boundary, then the splitting probability
$\mathcal{L}^{\mathrm{e}}\to 0$.  Using these facts in \eqref{gen} imposes $a_0=0$.
Together with the relations~\eqref{relations}, we finally obtain the
splitting probability to the left edge:
\begin{equation}\label{piasym}
\mathcal{L}^{\mathrm{e}}(y,\lambda)=e^{-\alpha y} +\sum\limits_{n=1}^{\infty} \big[
  e^{-(n+\alpha)y} - e^{ny}\big] e^{-n(n+\alpha)\lambda}\,.
\end{equation}
Note that the splitting probability to the right edge can be obtained from
\eqref{piasym} by 
$\mathcal{R}^{\mathrm{e}}(y,\lambda)\big|_{\alpha}=\mathcal{L}^{\mathrm{e}}(\lambda-y,\lambda)\big|_{1-\alpha}$.

To obtain the solution in the second formulation where an unbiased Brownian particle
starts at $x'_0=x_0\!-\!L_0/2$ in a cage whose boundaries are located at
$-L_0/2\!-\!c_1 t$ and $L_0/2\!+\!c_2 t$ at time $t$ (Fig.~\ref{setup}(b)),
we replace $\alpha$, $c$ and $x_0$ with their corresponding expressions in
terms of $c_1$, $c_2$ and $x'_0$ to give
\begin{align}
\label{asymforward}
\mathcal{L}^{\mathrm{e}}(x'_0,L_0)&=e^{-c_1(2x'_0+L_0)/2D}
 +   \sum\limits_{n=1}^{\infty} e^{-n[n(c_1+c_2)+c_1] L_0/D} \nonumber \\
& \qquad \qquad  \times \left\{
  e^{[-n(c_1+c_2)+c_1](2x'_0+L_0)/2D} - e^{[n(c_1+c_2)](2x'_0+L_0)/2D}\right\}\,.
\end{align}
When the initial interval length $L_0\gg 1$, the splitting probability is
well approximated by its first term
\begin{equation}\label{approxPi1}
\mathcal{L}^{\mathrm{e}}(x'_0,L_0) \sim e^{-c_1(2x'_0+L_0)/2D},
\end{equation}
which is exponentially larger than all higher-order terms in the series in
\eqref{asymforward}.

As a useful counterpoint, we can recover this last result by applying the
``free approximation''~\cite{Krapivsky:1996,Redner}, in which the
concentration within the interval is assumed to retain the same Gaussian form
as a diffusing particle on the infinite line with no imposed boundary
conditions. This approximation relies on the boundaries being outside the
range where the probability distribution is appreciable.  Thus we assume that
the concentration profile is
\begin{equation*}
c(x,t)=\frac{A(t)}{\sqrt{4\pi Dt}}\, e^{-(x-x'_0)^2/4Dt}\,,
\end{equation*}
where the unknown amplitude $A(t)$ accounts for the loss of probability
within the domain and should be determined self consistently.  In this free
approximation, the flux at any point in space is
\begin{equation*}
j= -D\frac{\partial c}{\partial x} = \frac{A(t)(x-x'_0)}{\sqrt{16 \pi Dt^3}}\,
  e^{-(x-x'_0)^2/4Dt}.
\end{equation*}
Thus the total flux leaving the cage is the sum of the flux at the two
boundaries:
\begin{equation*}
  \phi(t)=|j_1|+j_2= D \left. \frac{\partial c}{\partial x}\right|_{x=-\frac{L_0}{2}-c_1t} -D \left. \frac{\partial c}{\partial x}\right|_{x=\frac{L_0}{2}+c_2t}.
\end{equation*}
From the exiting flux, the rate equation for the overall amplitude $A(t)$ is
\begin{align}
\label{RELxv}
\frac{d A}{dt} = -A\left[ \frac{\frac{L_0}{2}-x'_0+c_2 t}{\sqrt{16\pi
      Dt^3}}\,e^{-(\frac{L_0}{2}-x'_0+c_2 t)^2/4Dt}
 +\frac{\frac{L_0}{2}+x'_0+c_1 t}{\sqrt{16\pi  Dt^3}}\,e^{-(\frac{L_0}{2}+x'_0+c_1 t)^2/4Dt}\right].
\end{align}
Integrating this equation to finite time, the  amplitude is
\begin{align}
\ln A(t)=-\frac{e^{-c_2(L_0-2x'_0)/2D}}{2}\,
\mathrm{erfc}\!\left(\frac{\frac{L_0}{2}\!-\!x'_0\!-\!c_2t}{\sqrt{4Dt}}\right) 
-\frac{e^{-c_1(L_0+2x'_0)/2D}}{2}\,
\mathrm{erfc}\!\left(\frac{\frac{L_0}{2}\!+\!x'_0\!-\!c_1  t}{\sqrt{4Dt}}\right).
\end{align}
For $t\to\infty$, this expression reduces to
\begin{align}
\label{Ainfv}
A(t\!=\!\infty)&=\exp\left[ -e^{-c_2(L_0-2 x'_0)/2D}- e^{-c_1(L_0+2x'_0)/2D}\right]\nonumber \\
& \sim 1-e^{-c_2(L_0-2 x'_0)/2D}- e^{-c_1(L_0+2x'_0)/2D}\qquad L_0\to\infty\,.
\end{align}
Here $A(t\!=\!\infty)$ represents the large-time limit of the survival
probability, namely, the probability that the particle has not touched either
of the boundaries.  It can thus be written as
$A(t\!=\!\infty)=1-\mathcal{L}^{\mathrm{e}}(x'_0,L_0)-\mathcal{R}^{\mathrm{e}}(x'_0,L_0)$,
which finally leads to
\begin{eqnarray}
\mathcal{L}^{\mathrm{e}}(x'_0,L_0) &\sim e^{-c_1 (L_0+2 x'_0)/2D}\qquad
\mathcal{R}^{\mathrm{e}}(x'_0,L_0) & \sim e^{-c_2(L_0-2 x'_0)/2D}\,,
\end{eqnarray}
in agreement with Eq.~\eqref{approxPi1}.

Unfortunately, this backward Fokker-Planck approach does not seem to be
adaptable to a contracting cage.  In this case, the general solution involves
a complex exponential dependence in $y$.  Therefore, the device used to
simplify the right boundary condition to the form given in \eqref{bc-right}
no longer holds.  To obtain the splitting probability in this case, we apply
a more general framework following the methods
of~\cite{Bray:2007b,Krapivsky:1996}.

\section{Contracting and Expanding Cage}
\label{sec:gen}

We now turn to the general case of a cage that can be either contracting or expanding.
Again, we study the problem in the reference frame of the left boundary that
is fixed at $x=0$.  For the expanding cage, the right boundary moves to the
right at speed $c$ and the particle drifts to the right with speed $\alpha c$
in addition to its diffusion.  For the contracting cage, the right boundary
and the particle both drift to the left.  Let $x$ denote the position of the
particle at time $t$.  We first compute the propagator $p(x,t)$ for the
particle in this cage, viz., the probability for the particle to be at $x$ at
time $t$, by solving the forward Fokker-Planck equation (see, e.g.,
\cite{Redner})
\begin{equation}\label{forward}
\partialderiv{p}{t} \pm \alpha c \frac{\partial p}{\partial x}=D \frac{\partial^2 p}{\partial x^2}\,,
\end{equation}
with the initial and boundary conditions $p(x,0)=\delta(x\!-\!x_0)$ and
$p(0,t)=p\big(L(t),t\big)=0$, and with $L(t)=L_0\pm ct$.  Here the upper sign
corresponds to the expanding cage and the lower sign to the contracting
cage throughout this section.  For the contracting cage the process stops
when the two boundaries meet at $t=L_0/c$.

When the boundaries are immobile, so that the cage length $L(t)=L_0$ is
constant, the elemental solutions of the Fokker-Planck equation with these
boundary conditions are well known~\cite{Redner}
\begin{equation*}
f_n(x,t)=\sin\left(\frac{n\pi x}{L_0}\right) \exp\left( \pm\frac{\alpha
    cx}{2D}-\frac{\alpha^2 c^2 t}{4D}  -\frac{n^2 \pi^2 Dt}{L_0^2} \right)\, \qquad \textrm{with $n\in\mathbb{N}$}.
\end{equation*}
To account for the interval length changing linearly with time, we follow the
method employed in~\cite{Krapivsky:1996} and adapted by Bray and
Smith~\cite{Bray:2007b} and postulate a solution to Eq.~\eqref{forward} with
absorbing boundary conditions at $x=0$ and $x=L(t)$ of the form
\begin{equation}
\label{pn}
p_n(x,t)=g(x,t) \sin\bigg( \frac{n\pi x}{L(t)}  \bigg) 
\exp \bigg(\pm\frac{\alpha cx}{2D}-\frac{\alpha^2 c^2 t}{4D} \bigg) 
\exp\bigg(-n^2 \pi^2D \int_0^t \frac{\diff{t'}}{L^2(t')} \bigg). 
\end{equation}
Substituting this trial function into Eq. \eqref{forward}, we obtain
\begin{equation}\label{gx}
\bigg( D \frac{\partial^2g}{\partial x^2} -\frac{\partial g}{\partial t}\bigg) 
\tan\bigg(\frac{n\pi x}{L(t)} \bigg)=-
\frac{n \pi}{L(t)} \bigg( 2D \frac{\partial g}{\partial x} \pm \frac{cx}{L(t)} g \bigg).
\end{equation}

We notice, as done in~\cite{Bray:2007b}, that we can seek a form for $g(x,t)$
that cancels both the left- and right-hand sides of \eqref{gx}.  Thus
$g(x,t)$ must simultaneously solve
\begin{align*}
\begin{split}
& D \frac{\partial^2g}{\partial x^2} =\frac{\partial g}{\partial t}\,, \\
& 2D \frac{\partial g}{\partial x}=\mp\frac{cx}{L(t)} g\,.
\end{split}
\end{align*}
These equations imply that the function $g(x,t)$ has the form
\begin{equation}
g(x,t)=\frac{K}{\sqrt{L(t)}}\,\, e^{\mp x^2 c/4D L(t)} \,, 
\end{equation}
with $K$ a constant. The general solution can now be written as a
superposition of the basis functions $p_n(x,t)$:
\begin{equation}
p(x,t)=  \sum\limits_{n\in\mathbb{N}} \frac{a_n}{\sqrt{L(t)}}
           \sin\bigg(\frac{n\pi x}{L(t)} \bigg)  
\exp\bigg(\mp\frac{x^2c}{4DL(t)} \pm\frac{\alpha c x}{2D}  \bigg)  
\exp\bigg(-\frac{\alpha^2 c^2 t}{4D}-\frac{n^2\pi^2 Dt}{L_0L(t)}\bigg)\,.
\end{equation}
For the initial condition $p(x,0)=\delta(x-x_0)$, we use the identity
\begin{equation*}
\sum\limits_{n\in\mathbb{N}} \sin\left(\frac{n\pi x}{L_0} \right) \sin\left(\frac{n\pi x_0}{L_0} \right) =\frac{L_0}{2} \,\delta(x-x_0),
\end{equation*}
for $0\leqslant x,x_0 \leqslant L_0$ to finally obtain the solution for the
given initial condition as
\begin{align}
p(x,t)&=  \sum\limits_{n\in\mathbb{N}} \frac{2}{\sqrt{L_0 L(t)}}
   \sin\bigg(\frac{n\pi x}{L(t)} \bigg)\sin\bigg(\frac{n\pi x_0}{L_0} \bigg)  \nonumber \\
& \qquad \times \exp\bigg(\mp\frac{c(x^2-x_0^2)}{4D L(t)} \pm\frac{\alpha c (x\!-\!x_0)}{2D} -\frac{\alpha^2 c^2 t}{4D}-\frac{n^2\pi^2 Dt}{L_0L(t)}\bigg)\,. 
\end{align}
The first-passage probability $F$ to the left edge of the cage is therefore
\begin{align}
F(0,t)&=D \partialderiv{p}{x}\Big|_{x=0}\,,\nonumber \\
&= \sum\limits_{n\in\mathbb{N}} \frac{2n\pi D}{\sqrt{L_0L^3(t)}}  \sin\!\left( \frac{n\pi x_0}{L_0}  \right) 
 \exp\!\bigg(\pm \frac{c x_0^2}{4DL_0} \mp\frac{\alpha c x_0}{2D} -\frac{\alpha^2 c^2 t}{4D}-\frac{n^2\pi^2 Dt}{L_0L(t)}\bigg)\,.
\end{align}

The splitting probability to the left edge is the time integral of this
first-passage probability.  As noted previously, the temporal integration
range depends on the sign of the speed.  Finally, the splitting
probabilities for the contracting and expanding cage, $\mathcal{L}^{\mathrm{c}}$
and $\mathcal{L}^{\mathrm{e}}$, respectively, are
\begin{subequations}
\label{pis}
\begin{align}\label{piplus}
\mathcal{L}^{\mathrm{c}}(x_0,L_0)=\int_0^{L_0/c} \diff{t} F(0,t) = &\sum\limits_{n\in\mathbb{N}} \frac{2n\pi D}{c\sqrt{L_0}} \sin\Big(
  \frac{n\pi x_0}{L_0} \Big) 
e^{-c(x_0-\alpha L_0)^2/4DL_0}  \nonumber \\
& \times  e^{n^2\pi^2 D/cL_0}\int_{0}^{L_0} \frac{\diff{L}}{L^{3/2}} \exp\Big(\frac{\alpha^2 cL}{4D}-\frac{n^2\pi^2 D}{cL}\Big), 
\end{align}
\begin{align}\label{pimoins}
\mathcal{L}^{\mathrm{e}}(x_0,L_0)=\int_0^{\infty} \diff{t} F(0,t) =& \sum\limits_{n\in\mathbb{N}} \frac{2n\pi D}{c\sqrt{L_0}} \sin\Big(
  \frac{n\pi x_0}{L_0} \Big) 
e^{c (x_0-\alpha L_0)^2/4DL_0}  \nonumber \\
& \times  e^{-n^2\pi^2 D/cL_0} \int_{L_0}^{\infty} \frac{\diff{L}}{L^{3/2}} \exp\Big(-\frac{\alpha^2 cL}{4D}+\frac{n^2\pi^2 D}{cL}\Big).
\end{align}
\end{subequations}
Expression \eqref{pimoins} for the splitting probability in an expanding cage
can be shown numerically to perfectly match the simpler form \eqref{piasym}
derived by the backward Fokker-Planck equation.  

However, these splitting probabilities are not convenient for numerical
evaluation.  Instead, it is expedient to use the Poisson summation
formula~\cite{Olver}
\begin{equation*}
\sum_{n\in \mathbb{Z}} h(n)= \sum_{m\in\mathbb{Z}} \hat{h}(2\pi m),
\end{equation*}
with $\hat{h}(x)=\int_{-\infty}^{+\infty} \diff{t} e^{-ixt} f(t)$, to give
the alternative expression 
\begin{align}
  &\mathcal{L}^{\mathrm{c}}(x_0,L_0)= \sum\limits_{m\in\mathbb{Z}} \sqrt{\frac{c}{4\pi D}}
    e^{-c(x_0-\alpha L_0)^2/4DL_0}\!
\int_0^{L_0} \frac{\diff{L}}{(L_0\!-\!L)^{3/2}} \; e^{\alpha^2 cL/4D} \nonumber \\
  & \quad \times \!\exp\!\bigg[\!-\frac{cL(4L_0^2m^2\!+\!x_0^2)}{4DL_0(L_0\!-\!L)}\bigg]  
\left\{x_0 \cosh\bigg[\frac{cLx_0m}{D(L_0\!-\!L)}\bigg] 
- 2mL_0 \sinh\bigg[ \frac{cLx_0m}{D(L_0\!-\!L)}\bigg] \right\},
\end{align}
which is more suitable for numerical evaluation (and similarly for $\mathcal{L}^{\mathrm{e}}(x_0,L_0)$).

\section{Optimal Capture Criterion}
\label{sec:disc}

We now turn to our original question: what is the optimal strategy for the
shepherd to catch her skittish lamb without driving it over the precipice?
In the shepherd's reference frame, the lamb approaches at speed $\alpha c$
while the precipice approaches at a higher speed $c$.  The probability to
catch the lamb in this contracting cage is the splitting probability
$\mathcal{L}^{\mathrm{c}}(x_0,L_0)$ \eqref{piplus}.  What speed $c$ maximizes this
splitting probability?

Partial conclusions can be drawn by studying the limits of $c\to 0$ and
$c\to\infty$.  If $c=0$, the splitting probability
$\mathcal{L}^{\mathrm{c}}(x_0,L_0)$ is a linear function of its initial position
(see, e.g.,~\cite{Redner})
\begin{equation*}
\mathcal{L}^{\mathrm{c}}(x_0,L_0)=\frac{L_0-x_0}{L_0}.
\end{equation*}
The qualitative behavior of $\mathcal{L}^{\mathrm{c}}(x_0,L_0)$ when $c\to\infty$ can also be
easily understood.  In this limit, if the time $x_0/(\alpha c)$ for the lamb
to reach the shepherd is smaller than the time $(L_0-x_0)/[(1-\alpha)c]$ for
the precipice to catch up to the lamb, then $\mathcal{L}^{\mathrm{c}}(x_0,L_0)\to 1$, while
$\mathcal{L}^{\mathrm{c}}(x_0,L_0)\to 0$ otherwise.  These two times match when
$\alpha=x_0/L$.  Thus in the limit of large speed, the splitting probability
reduces to a step function, with $\mathcal{L}^{\mathrm{c}}(x_0,L_0)\approx 1$ for
$x_0/L<\alpha$ and $\mathcal{L}^{\mathrm{c}}(x_0,L_0)\approx 0$ for $x_0/L>\alpha$ .

Let us now focus on the first order of the splitting probability for small speeds.  In this
case, the integral over $L$ in Eq.~\eqref{piplus}
\begin{equation*}
\mathcal{I} \equiv \int_{0}^{L_0} \frac{\diff{L}}{L^{3/2}} \exp\left(\frac{\alpha^2 cL}{4D}-\frac{n^2\pi^2 D}{cL}\right)
\end{equation*}
can be developed with respect to $c$ by expanding
$\exp\left(\alpha^2 c L/4D\right)$ and integrating by parts.  This yields
\begin{equation}
\mathcal{I} \sim \exp\left(-\frac{n^2\pi^2D}{ c L_0}\right) \left[ \frac{c\sqrt{L_0}}{n^2\pi^2D} -\frac{c^2L_0^{3/2}}{2n^4\pi^4D^2} \left(1-\frac{n^2\pi^2\alpha^2}{2} \right) \right].
\end{equation}
Then using
\begin{align}
&\sum\limits_{n=1}^{\infty} \frac{(-1)^n}{n} \sin(nz)=-\frac{z}{2}\,, \nonumber\\
&\sum\limits_{n=1}^{\infty} \frac{(-1)^n}{n^3} \sin(nz)= \frac{z^3}{12} -\frac{\pi^2 z}{12}\,,\nonumber
\end{align}
we obtain the small-speed behavior of the splitting probability:
\begin{equation}
\label{Pi+-smallc}
\mathcal{L}^{\mathrm{e,c}}(x_0,L_0) = \frac{L_0-x_0}{L_0} \mp \frac{cx_0}{6D}  \frac{(L_0-x_0) \big[(3\alpha-1)L_0-x_0\big]}{L_0^2}\, +o(c).
\end{equation}
Here we also quote the limiting form splitting probability for the expanding
cage, which is found from \eqref{pimoins} by the same steps as outlined above.

\begin{figure}[ht]
\centerline{\qquad\qquad
\includegraphics[width=180pt]{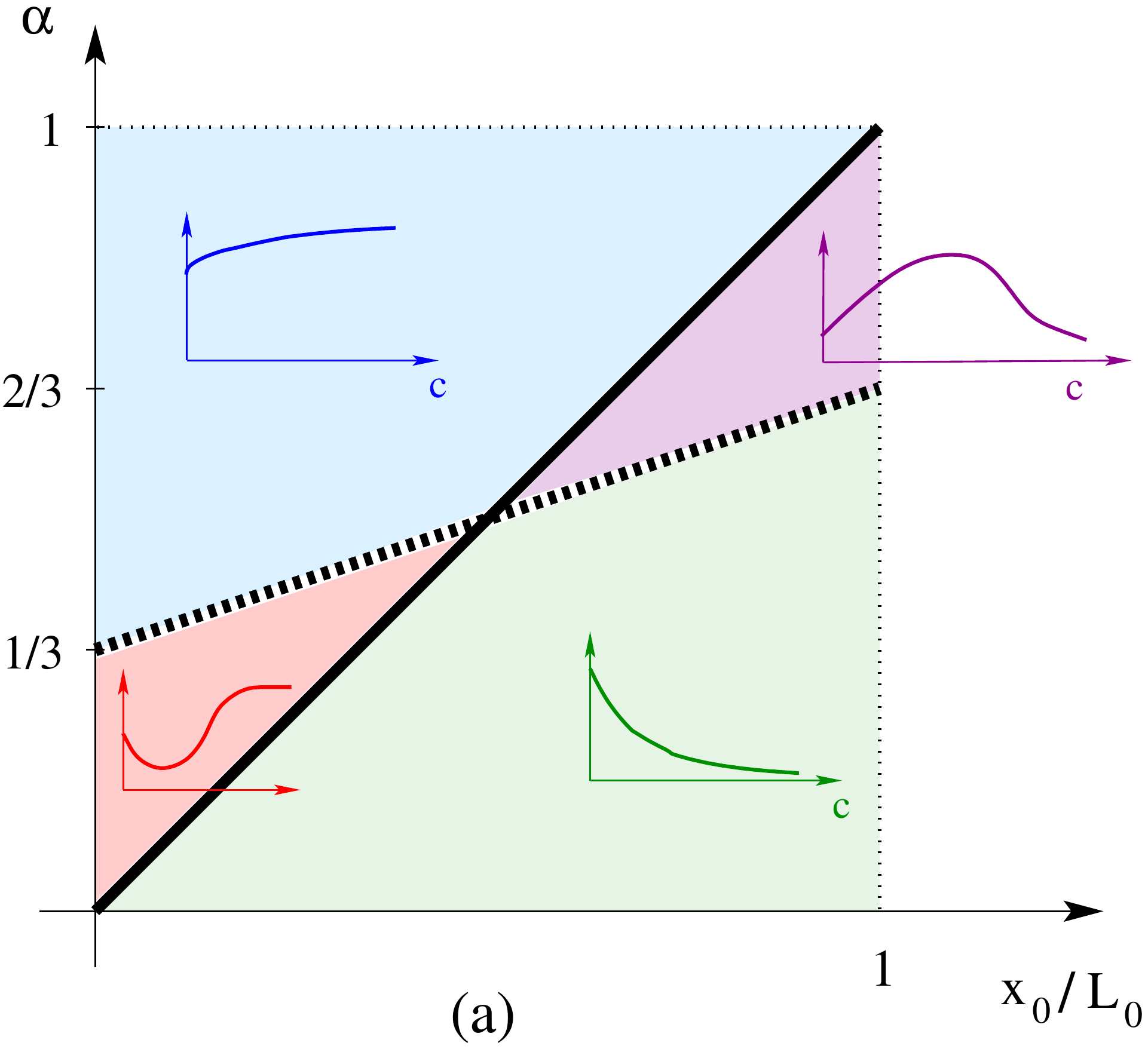}\qquad\qquad
\includegraphics[width=180pt]{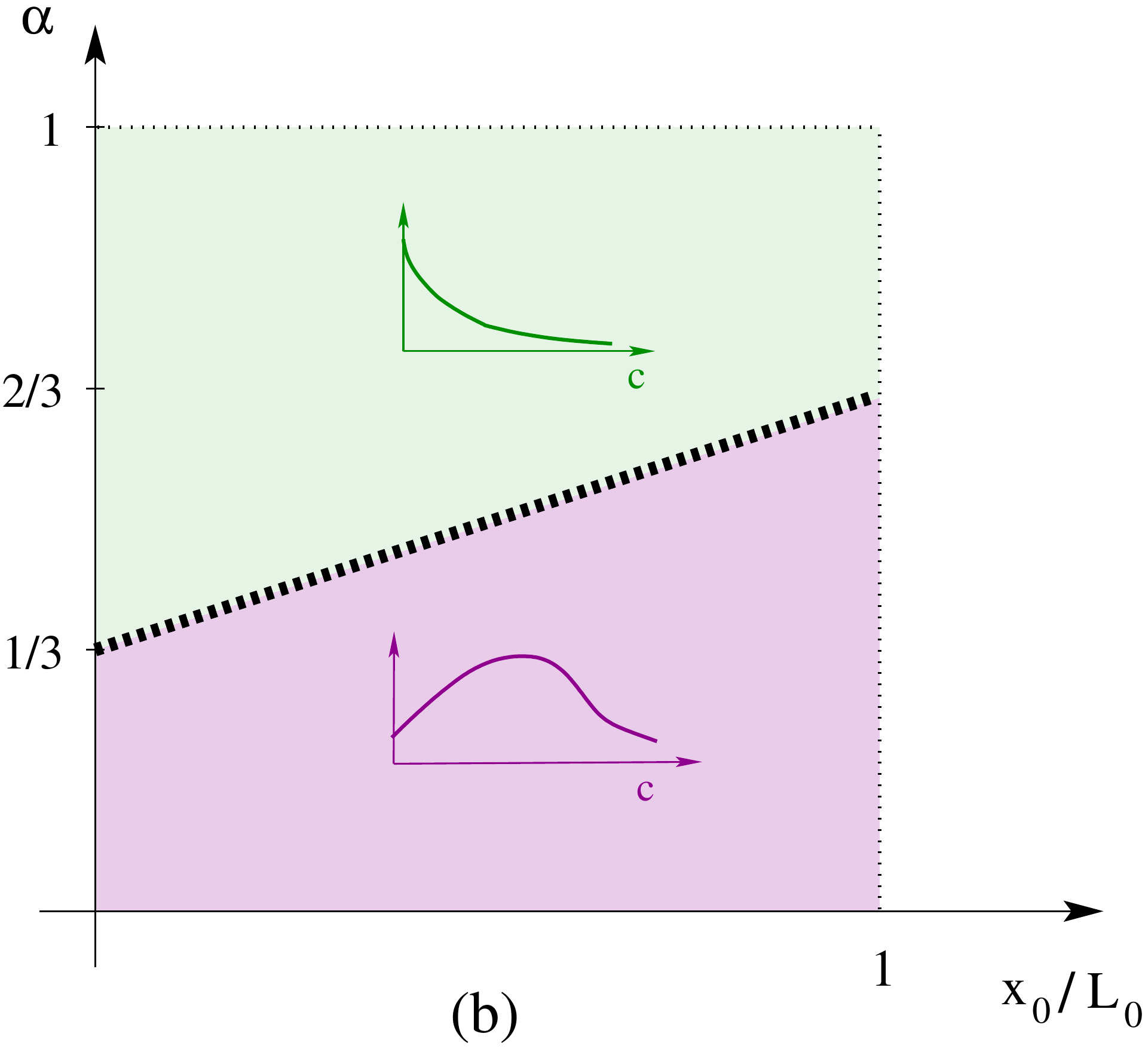}}
\caption{Splitting probability phase diagram for the (a) contracting and (b)
  expanding cage in the speed ratio ($\alpha$) and initial position ($x_0$)
  plane.  In (a), $\mathcal{L}^{\mathrm{c}}\approx 1$ for $c\to\infty$ above the
  solid line, while $\mathcal{L}^{\mathrm{c}}\approx 0$ for $c\to\infty$ below.
  Above the dashed line $\mathcal{L}^{\mathrm{c}}$ is an increasing function of $c$
  at $c=0$ and a decreasing function below.  These lines delineate four zones
  of behavior as discussed in the text.  In (b), there are two zones.}
\label{diagphase1}
\end{figure}

As a result of the $\alpha$ and $x_0$ dependence of the first-order term in
$c$ given above, $\mathcal{L}^{\mathrm{c}}(x_0,L_0)$ is an increasing function of
speed at $c=0$ when $\alpha>\frac{1}{3}(x_0/L_0\!+\!1)$ and decreasing otherwise.
Combining this fact with the limiting behavior for $c\to\infty$, we deduce
that the splitting probability can be a non-monotonic function of the speed,
for specific values of $x_0$ and $\alpha$.  This leads to rich behaviors for
the splitting probability, as illustrated in Fig.~\ref{diagphase1}(a).

We can now give advice to the shepherd.  There are four distinct strategies,
corresponding to the four zones of Fig.~\ref{diagphase1}(a):
\begin{enumerate}

\item ``Dangerous zone'' (lower right).  Here, either the lamb is very
  fearful, $\alpha\ll 1$, or sufficiently close to the precipice,
  $\alpha<\frac{1}{3}(x_0/L_0\!+\!1)$ and $\alpha<x_0/L_0$, so that
  $\mathcal{L}^{\mathrm{c}}$ monotonically decreases with $c$.  Thus the optimum
  strategy for the shepherd is to not move.

\item ``Safe zone'' (upper left).  Here, either the lamb is not very fearful
  or is sufficiently close to the shepherd, $\alpha>\frac{1}{3}(x_0/L_0\!+\!1)$
  and $\alpha>x_0/L_0$, so that $\mathcal{L}^{\mathrm{c}}$ monotonically increases
  with $c$.  The shepherd should run as fast as possible to maximize the
  probability to catch the lamb.

\item ``Optimizable zone'' (upper right).  Here, the lamb is close to the
  precipice but not too fearful, $\alpha>\frac{1}{3}(x_0/L_0\!+\!1)$ and
  $\alpha<x_0/L_0$ so that $\mathcal{L}^{\mathrm{c}}$ has a maximum as a function of $c$.  Thus
  there is an optimal speed that maximizes the probability for the shepherd
  to catch the lamb.

\item ``Dilemma zone'' (lower left).  Here the lamb is close to the shepherd
  but also very fearful, $\alpha<\frac{1}{3}(x_0/L_0\!+\!1)$ and
  $\alpha>x_0/L_0$.  Thus $\mathcal{L}^{\mathrm{c}}$ initially decreases with $c$
  before eventually increasing.  Thus if the shepherd is unfit, she should
  stay still.  However, if she is sufficiently fit, she should run as fast as
  possible.
\end{enumerate}

For the expanding cage, the splitting probability given in
Eq.~\eqref{Pi+-smallc} is now an increasing function of speed at $c=0$ for
$\alpha<\frac{1}{3}(x_0/L_0\!+\!1)$.  Using this small-speed dependence, together
with the limiting behavior $\mathcal{L}^{\mathrm{e}}(x_0,L_0)\to 0$ for large speed,
gives the phase diagram shown Fig.~\ref{diagphase1}(b).  There again exists a
zone in the phase diagram where the splitting probability can be maximized
with respect to the speed.  Here, the maximum of the splitting probability
arises from the interplay between two competing effects from the cage
expansion.  Indeed, consider the complementary probability that the lamb
never reaches the left boundary.  The larger the speed, the lower the
probability for the lamb to reach the left boundary.  At the same time, the
larger the speed, the higher the probability for the lamb to also \emph{not}
reach the right boundary, which implicitly increases the probability to reach
the left boundary.  As a result of these competing effects, there exists an
optimal speed of expansion that maximizes the splitting probability.

\begin{figure}[h]
\centering
\includegraphics[width=180pt]{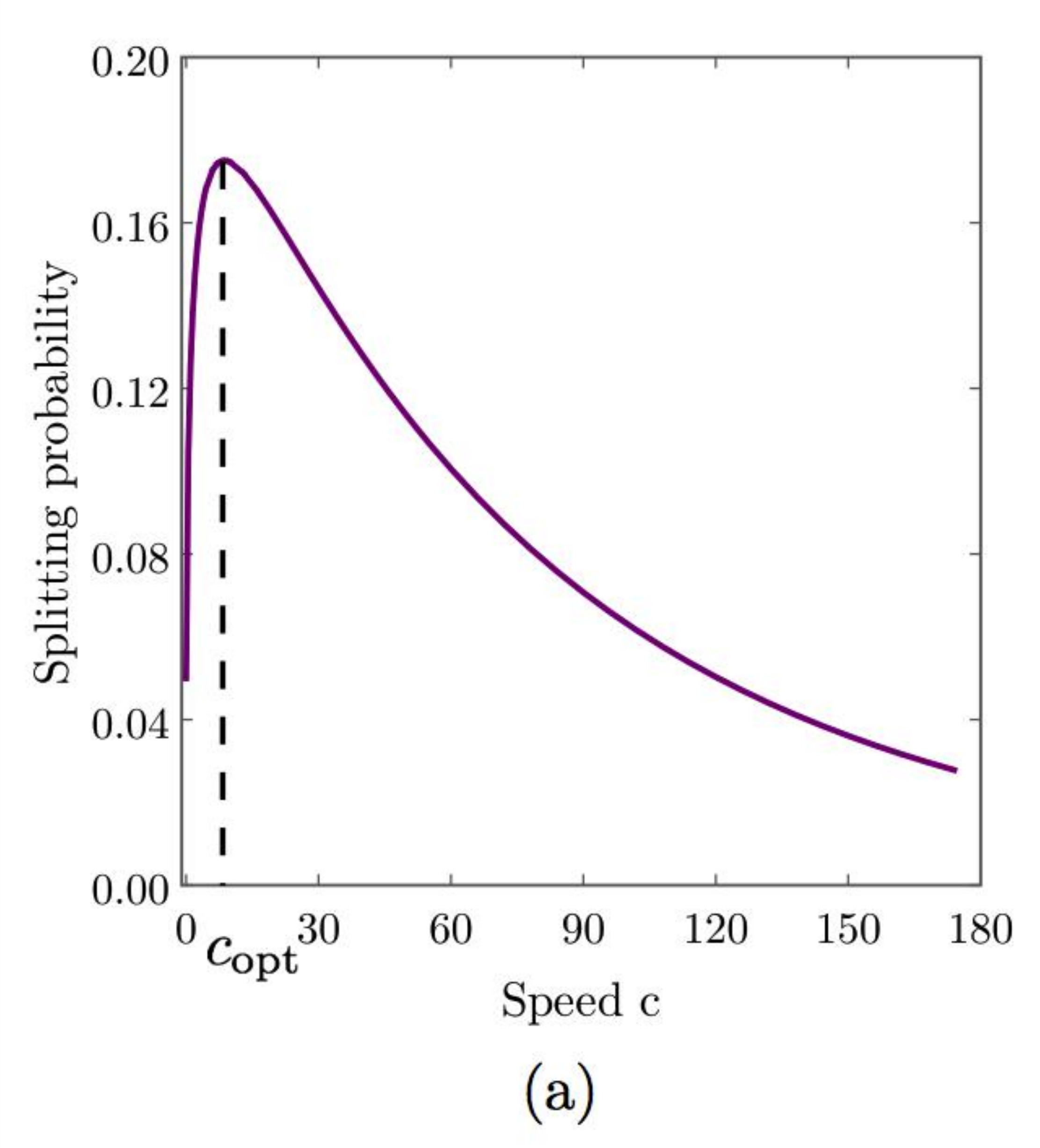}\qquad\qquad
\includegraphics[width=180pt]{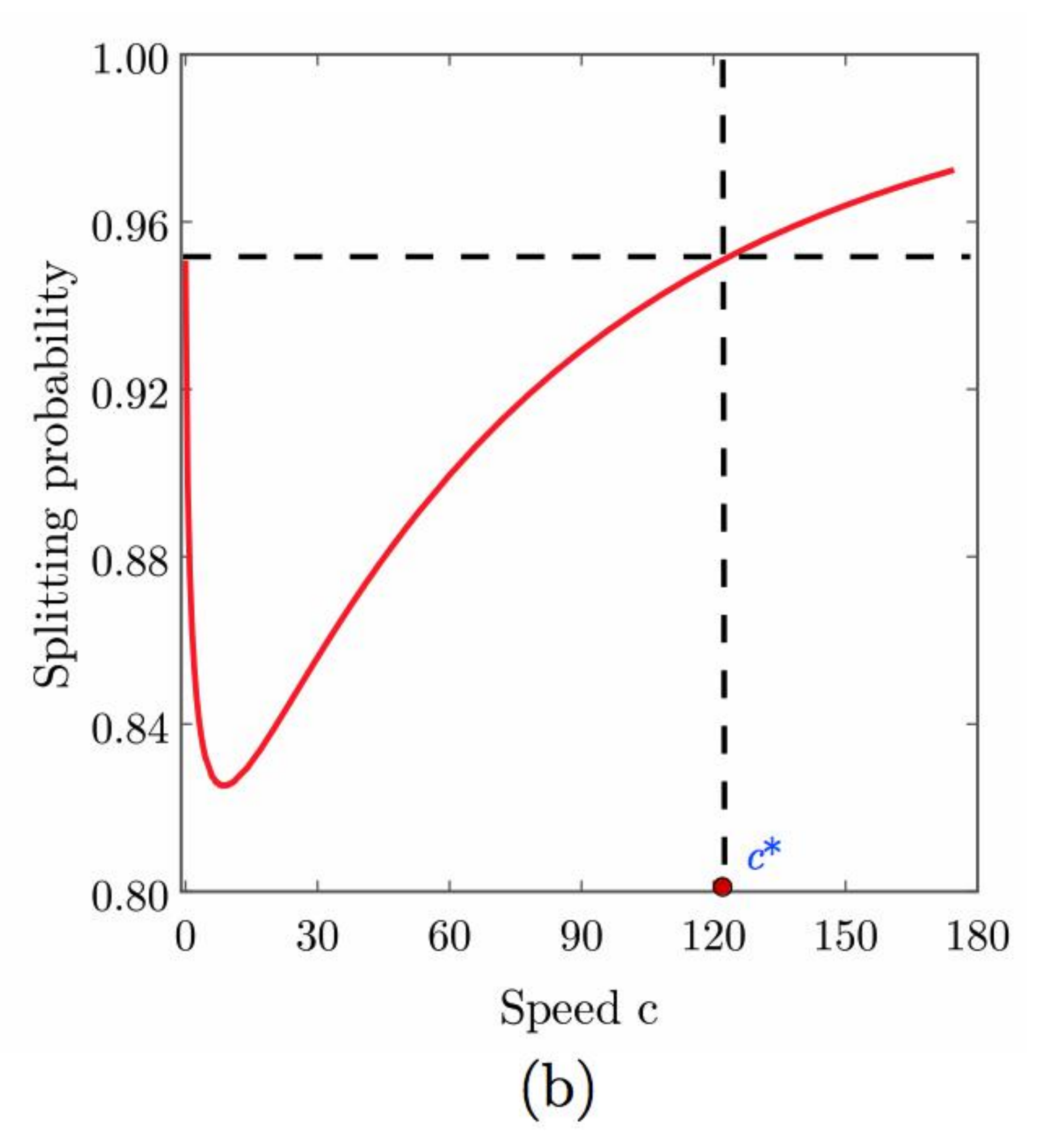} 
\caption{Splitting probability to catch the lamb in the (a) ``optimizable
  zone'' and (b) ``dilemma zone''.  In (a), the splitting probability is
  increased substantially (by more than a factor 3 for $\alpha=0.92$,
  $x_0=19$, and $L_0=20$) when the shepherd moves at the optimal speed
  $c_{\textrm{opt}}$ instead of staying still.  In (b), the probability to
  catch the lamb is diminished if the shepherd runs slower than speed $c^*$.
  Here the parameter values are $\alpha=0.08$, $x_0=1$ and $L_0=20$).}
\label{minmax}
\end{figure}

\section{Conclusion}

We analytically determined the splitting probabilities for a one-dimensional
Brownian motion in a cage whose boundaries move at constant speeds $c_1$ and
$c_2$.  We analyzed both the cases of contracting and expanding cages.  In
addition, we calculated the time-dependent first-passage probabilities at
each of the boundaries.  Intriguing behaviors of the splitting probabilities
arise as a consequence of the ballistic boundary motion.  Indeed, we found that the
splitting probabilities can vary non-monotonically with the relative speeds
of the boundaries, depending on the initial position of the Brownian particle
and the ratio between the two boundary speeds.  In the context of a fearful
lamb near a precipice (Fig.~\ref{setup}), this non-monotonicity defines a
non-trivial optimal strategy for the shepherd to catch the lamb.

This approach could be extended to determine the splitting probability to a
subset of a growing $d$-dimensional sphere for a particle that starts
somewhere within the interior.  It would also be interesting to extend the
conditional exit problem to the case of non-linear displacements of the
boundaries where
the lamb capture probability should also have a non-trivial optimization.

\end{document}